\documentclass[aps,twocolumn,amsmath,amssymb,showpacs,prl,superscriptaddress,unsortedaddress]{revtex4}
\usepackage{epsf}
\usepackage{graphicx}
\usepackage{gensymb}

\newcommand{\etal}{{\it et al.}}

\begin{document}

\title{Three- to Two-Dimensional Transition of the Electronic Structure in CaFe$_2$As$_2$ - parent compound for an iron arsenic high temperature superconductor}

\author{Chang Liu}
\affiliation{Ames Laboratory and Department of Physics and
Astronomy, Iowa State University, Ames, Iowa 50011, USA}

\author{Takeshi Kondo}
\affiliation{Ames Laboratory and Department of Physics and
Astronomy, Iowa State University, Ames, Iowa 50011, USA}

\author{Ni Ni}
\affiliation{Ames Laboratory and Department of Physics and
Astronomy, Iowa State University, Ames, Iowa 50011, USA}

\author{A. D. Palczewski}
\affiliation{Ames Laboratory and Department of Physics and
Astronomy, Iowa State University, Ames, Iowa 50011, USA}

\author{A. Bostwick}
\affiliation{Advanced Light Source, Berkeley National Laboratory,
Berkeley, California 94720, USA}

\author{G. D.~Samolyuk}
\affiliation{Ames Laboratory and Department of Physics and
Astronomy, Iowa State University, Ames, Iowa 50011, USA}

\author{R.~Khasanov}
\affiliation{Laboratory for Muon Spin Spectroscopy,
Paul Scherrer Institut, CH-5232 Villigen PSI, Switzerland}

\author{M.~Shi}
\affiliation{Swiss Light Source, Paul Scherrer Institute, 5232
Villingen PSI, Switzerland}

\author{E. Rotenberg}
\affiliation{Advanced Light Source, Berkeley National Laboratory,
Berkeley, California 94720, USA}

\author{S. L. Bud'ko}
\affiliation{Ames Laboratory and Department of Physics and
Astronomy, Iowa State University, Ames, Iowa 50011, USA}

\author{P. C. Canfield}
\affiliation{Ames Laboratory and Department of Physics and
Astronomy, Iowa State University, Ames, Iowa 50011, USA}

\author{A. Kaminski}
\affiliation{Ames Laboratory and Department of Physics and
Astronomy, Iowa State University, Ames, Iowa 50011, USA}

\date{\today}
\begin{abstract}
We use angle-resolved photoemission spectroscopy (ARPES) to study
the electronic properties of CaFe$_2$As$_2$ - parent compound of a
pnictide superconductor. We find that the structural and magnetic
transition is accompanied by a three- to two-dimensional (3D-2D)
crossover in the electronic structure. Above the transition
temperature ($T_s$) Fermi surfaces around $\Gamma$ and X points are
cylindrical and quasi-2D. Below $T_s$ the former becomes a 3D
ellipsoid, while the latter remains quasi-2D. This finding strongly
suggests that low dimensionality plays an important role in
understanding the superconducting mechanism in pnictides.
\end{abstract}


\maketitle

The dimensionality of electronic structure plays an important role
in the superconductivity of solids. The cuprate superconductors have
the highest known transition temperatures and they have quasi two
dimensional (2D) electronic structure. In contrast, the borocarbides
\cite{Gupta_B2C},
%%%%%%%%%%%%% revised
another class of relatively high transition
temperature superconductors,
%%%%%%%%%%%%%
have a strictly three dimensional (3D) electronic structure
\cite{Starowicz}. The situation in the newly discovered iron arsenic
high temperature superconductors is less clear. Both $R$FeAsO
(R1111, $R$ being the rare earth element) \cite{Original} and
$A$Fe$_2$As$_2$ (A122, $A$ being Ca, Sr, Ba) \cite{Rotter} have
strongly layered structures, with the iron-arsenic layers believed
to be mainly responsible for the electronic properties and
superconductivity.
%%%%%%%%%Added
The A122 family share the same ThCr$_2$Si$_2$ structure as the
borocarbides and both display interesting interplay between
magnetism and superconductivity.
%%%%%%%%%
Some band structure calculations predict strong $k_z$ dispersion and
3D Fermi surfaces (FSs) \cite{FengjieMa} in the magnetic state of
the parent compounds, however so far this has not been observed by
angle-resolved photoemission spectroscopy (ARPES)
\cite{ChangLiu,TakeshiPRL,Ding_Europhys,Zhou,Wray,Hsieh,Ding_2}. On
the other hand, a number of physical properties display an
anisotropy that is a few orders of magnitude smaller than that found
in the cuprates \cite{NiNi_Hc2,Martin,Tanatar}, which hints at a
strong hopping between the layers. In terms of the pairing
mechanism, most theoretical models depend on nesting between the
$\Gamma$- and X(M)-pockets \cite{Mazin,Dong}, for which the
dimensionality of the FSs is very important.
%Penetration depth measurements report node in the superconducting gap \cite{Ruslan1, Ruslan2}, while ARPES measurement performed for particular value of kz momentum found isotropic value of the superconducting gap.
%REMOVED
%Interestingly penetration depth measurements demonstrate that in the
%R1111 family the superconducting (SC) gap is nodeless
%\cite{Martin,Malone}, while the data in the A122 system is
%consistent with the presence of nodes \cite{Ruslan1, Ruslan2}, yet
%no such nodes have been observed in previous ARPES measurements
%\cite{TakeshiPRL,Ding_Europhys,Zhou,Wray} which probe only a single
%$k_z$.
%%%%%%%%%%Added
%Whether there is gap node(s) located at other $k_z$s is therefore
%still an open question.
%%%%%%%%%%

The parent compounds of the two iron pnictide families undergo a
tetragonal-to-orthorhombic (\textsf{Tet}-\textsf{Ortho})
structural/magnetic transition at elevated temperatures
\cite{Nomura,NiNi_Hc2}. Superconductivity appears in the A122
materials when the high temperature \textsf{Tet}-phase is stabilized
down to low enough temperatures (usually by doping with K in a Ba
position or Co substitution for Fe which lowers the transition
temperature) \cite{NiNiCo}. Determining the differences between
these phases is key to gaining a full understanding and control of
the superconductivity in the iron-pnictides. In particular,
CaFe$_2$As$_2$ shows a first order \textsf{Tet}-\textsf{Ortho}
transition with a transition temperature (160 $\sim$ 170K) depending
on the growth method \cite{NiNi_transition,Kumar,Wu_FeAsflux}.
Superconductivity is found when moderate (possibly not purely
hydrostatic) pressure is applied \cite{debate}, or by doping with
cobalt \cite{Kumar} or sodium \cite{Wu_FeAsflux}. Such a rich phase
diagram makes CaFe$_2$As$_2$ an ideal system for studying the
electronic properties of the parent compounds. In this Letter we
present ARPES data which demonstrates that the electronic structure
in the parent compound of the A122 iron arsenic family undergoes a
3D to 2D crossover associated with the structural/magnetic
transition. The cylindrical, quasi-2D FS surrounding the
$\Gamma$-point becomes highly dispersive (3D) and forms an ellipsoid
upon cooling below the transition temperature into the
\textsf{Ortho}-phase. The observation of the 3D FS is consistent
with recent results from quantum oscillation measurements
\cite{Sebastian, Harrison}. The FSs around the X-pocket remain
quasi-2D and unaffected by the structural transition.

\begin{figure}
\includegraphics[width=2.8in]{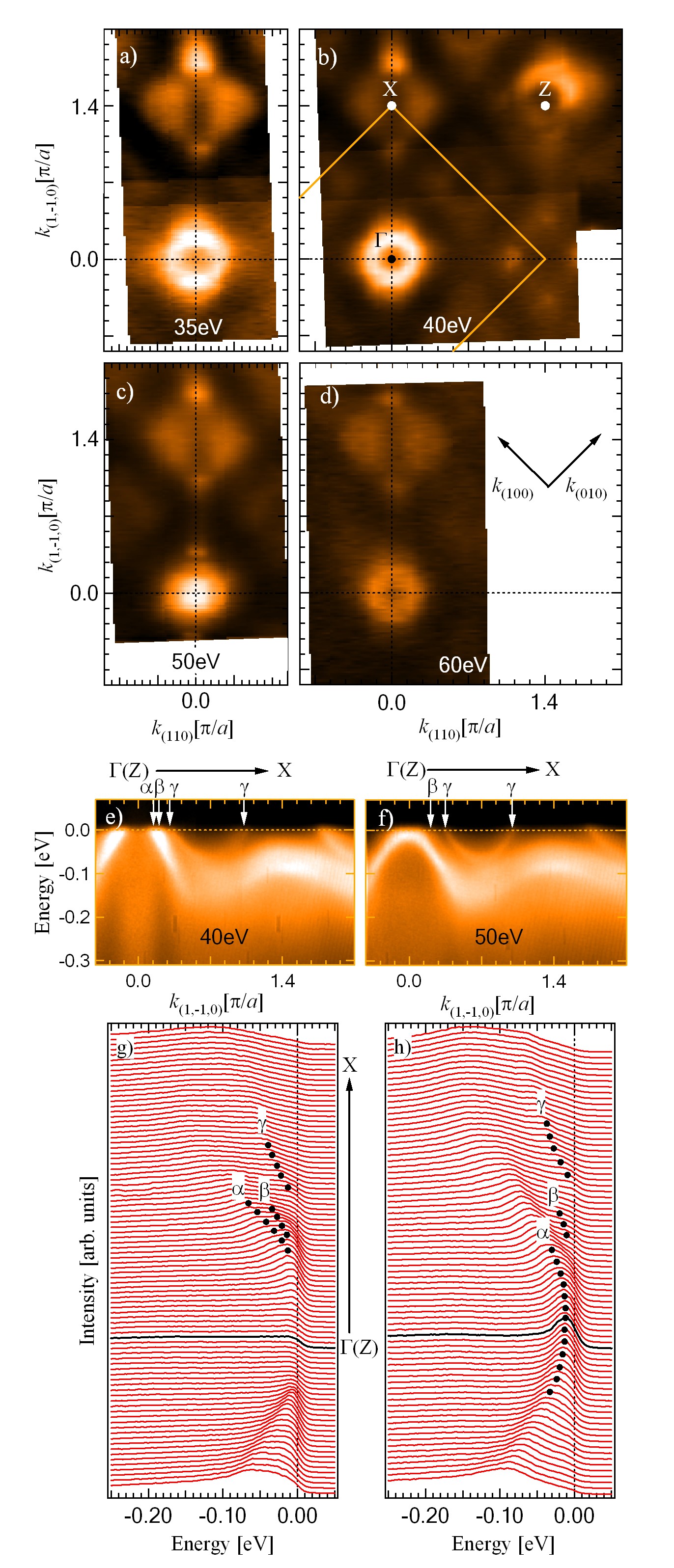}
\caption{(color online) Fermi surface maps of CaFe$_2$As$_2$ low
temperature orthorhombic phase ($T=12$K) for a few photon
energies. (a)-(d) ARPES intensity integrated within 10 meV about $\mu$
 for $h\nu=35,40,50$ and 60eV, respectively. Bright areas mark the location of the Fermi
surfaces. (e)-(f) Band dispersion data along the $\Gamma$-X
direction for $h\nu=40$ and 50eV. $\alpha, \beta,
\gamma$ correspond to different bands that cross $\mu$. (g)-(h)
Energy distribution curves (EDCs) for data in panels (e)-(f) over
the same $k$ range.} \label{fig1}
\end{figure}

Plate like single crystals of CaFe$_2$As$_2$ were grown out of a FeAs flux as
well as Sn flux using conventional high-temperature solution growth
techniques \cite{NiNi_transition} with typical dimensions ranging from 2 $\times$ 2 mm$^2$ up to 10
$\times$ 10 mm$^2$. Following the growth the samples were annealed
at 500$^\circ$C for 24 hours. Resistivity measurements showed a
first-order \textsf{Tet-Ortho} transition at $\sim$160K for the FeAs
flux grown samples and $\sim$170K when using Sn flux.
%For FeAs flux growth, small Ca
%chunks, FeAs powder were mixed together according to the ratio Ca :
%FeAs = 1:4. The mixture was placed into an alumina crucible. A
%second catch crucible containing quartz wool was placed on top of
%this growth crucible and both were sealed in a quartz tube under 1/3
%atmosphere Ar gas. The sealed quartz tube was heated up to
%1180$^\circ$C, stayed there for 2 hours, fast cooled to
%1020$^\circ$C, and then slowly cooled to 960$^\circ$C over 36 hours
%before the FeAs was decanted by a centrifuge, yielding plate-like
%
 The ARPES measurements are performed at beam lines 10.0.1 and 7.0.1 of the
Advanced Light Source (ALS), Berkeley, California, and the SIS beam
line of the Swiss Light Source, Switzerland. Energy resolution was
set at $20\sim30$meV, vacuum conditions were better than
$5\times10^{-11}$ torr. All samples were cleaved \textit{in situ}
along the (001) plane, yielding mirror-like, clean surfaces. Lattice
constant values from Ref. \cite{Goldman_lattice} are used to
determine the $k$-space positions. The high symmetry points X and Z
for both two phases are defined to be ($\pi/a$, $\pi/a$($b$), 0) and
(0, 0, $2\pi/c$), respectively, with $k_x$ ($k_{(100)}$) and $k_y$
($k_{(010)}$) axes along the Fe-As bonds.

FS maps of CaFe$_2$As$_2$ obtained at several different photon
energies in the \textsf{Ortho}-phase are shown in Fig. 1 along with
ARPES intensity plots and energy distribution curves (EDCs). Varying
the photon energy in ARPES effectively changes the momentum offset
along the direction perpendicular to the sample surface (in our case
this direction corresponds to $k_z$ - perpendicular to the Fe-As
layers) \cite{Hufner,TakeshiThesis}. The striking feature of Fig. 1
is that the dispersion of one of the bands that form the
``$\Gamma$-pocket" (Fermi contour around the zone center
$k_x=k_y=0$) changes dramatically with photon energy. At
$h\nu=40$eV, three different bands $\alpha, \beta$ and $\gamma$
cross the chemical potential ($\mu$) as seen in Figs. 1(e), 1(g) and
this gives rise to three FS sheets. The $\gamma$-band crosses $\mu$
again near the X-point, forming a characteristic flower shape of the
X-pocket. The other two bands ($\alpha, \beta$) are relatively close
to each other. They form two hole pockets around $\Gamma$. Similar
behavior has been reported in Ba$_{1-x}$K$_x$Fe$_2$As$_2$
\cite{Feng_Ba,Ding_Europhys,Zhou,Sato_K,Borisenko}. However, at
$h\nu=50$eV, the top of the $\alpha$ band is located below $\mu$ and
the corresponding $\alpha$-pocket disappears. The $\beta$- and
$\gamma$-pockets continue to cross $\mu$ at about the same
$k_\textrm{F}$ as for 40eV photons [Figs. 1(c), 1(f) and 1(h)].
These data conclusively demonstrate that the electronic structure in
the \textsf{Ortho}-phase of CaFe$_2$As$_2$ is 3D. On the other hand,
no obvious change is seen for the X-pocket at these 4 energies.

A comprehensive study of the evolution of the $\Gamma$- and X-pocket
with different incident photon energies is shown in Fig. 2 for the
\textsf{Ortho}-phase of CaFe$_2$As$_2$. The photon energy ranges
from 35 to 105eV. The FS map along the $k_z$ direction is shown in
Fig. 2(a) over a range corresponding to several Brillouin zones. The
$\alpha$ Fermi pocket forms an ellipsoid centered at $\Gamma$ in the
respective Brillouin zones with a $4\pi/c$ periodicity. It should be
noted that the observation of $k_z$ dispersion with such periodicity
clearly demonstrates that ARPES data from these samples reveals
intrinsic, bulk electronic properties. In Fig. 2(b) we extract the
Fermi crossing momenta ($k_\textrm{F}$s) from the momentum
distribution curve (MDC) peaks at $\mu$ for each photon energy. It
is clear that only the $\alpha$-band (black solid dots) but not the
$\beta$- and $\gamma$-bands show a strong $k_z$ dispersion. Almost
no dispersion of the X-pocket is observed, which indicates its quasi
2D nature. The consistency of this map with that of Fig. 1 is better
seen at Figs 2(c)-2(e) where ARPES intensity maps are shown for 3
high symmetry points. At $h\nu=58$eV ($k_z=16\pi$) the $\alpha$-band
does not cross $\mu$, while at $h\nu=80$ and 41eV ($k_z=18\pi$ and
$14\pi$) all three bands form Fermi pockets.

\begin{figure}
\includegraphics[width=3.2in]{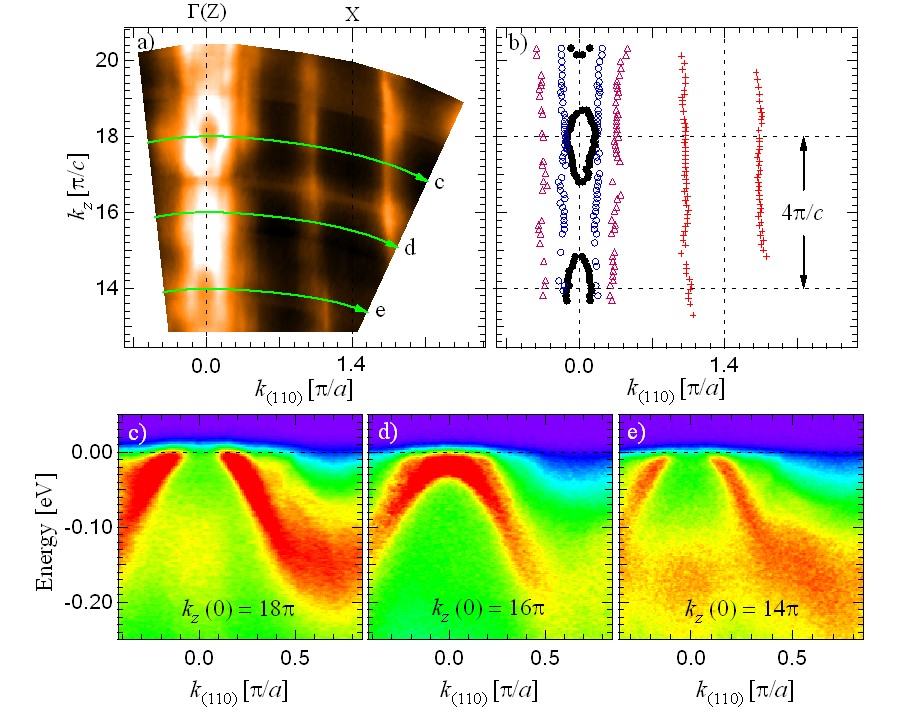}
\caption{(color online) $k_z$ dispersion data for CaFe$_2$As$_2$ low
temperature orthorhombic phase ($T=40$K). (a) $k_z$ dispersion data
obtained by plotting ARPES intensity integrated within 10 meV about
$\mu$ as a function of $k_{(110)}$ and the energy of the incident
photons (which corresponds to different values of $k_z$). The photon
energy range used was from 35 to 105 eV. (b) Fermi momentum
$k_\textrm{F}$ was extracted from the data in panel (a) using the
peak positions of the MDCs. (c)-(e) Band dispersion
data along the $\Gamma$-X direction for $k_z=18, 16$ and $14\pi$ at
$k_{(110)}=0$ (corresponding to photon energies of
$h\nu=80, 58$ and 41eV, respectively). The locations of these cuts are marked by green lines in panel
(a). } \label{fig2}
\end{figure}

\begin{figure}
\includegraphics[width=2.9in]{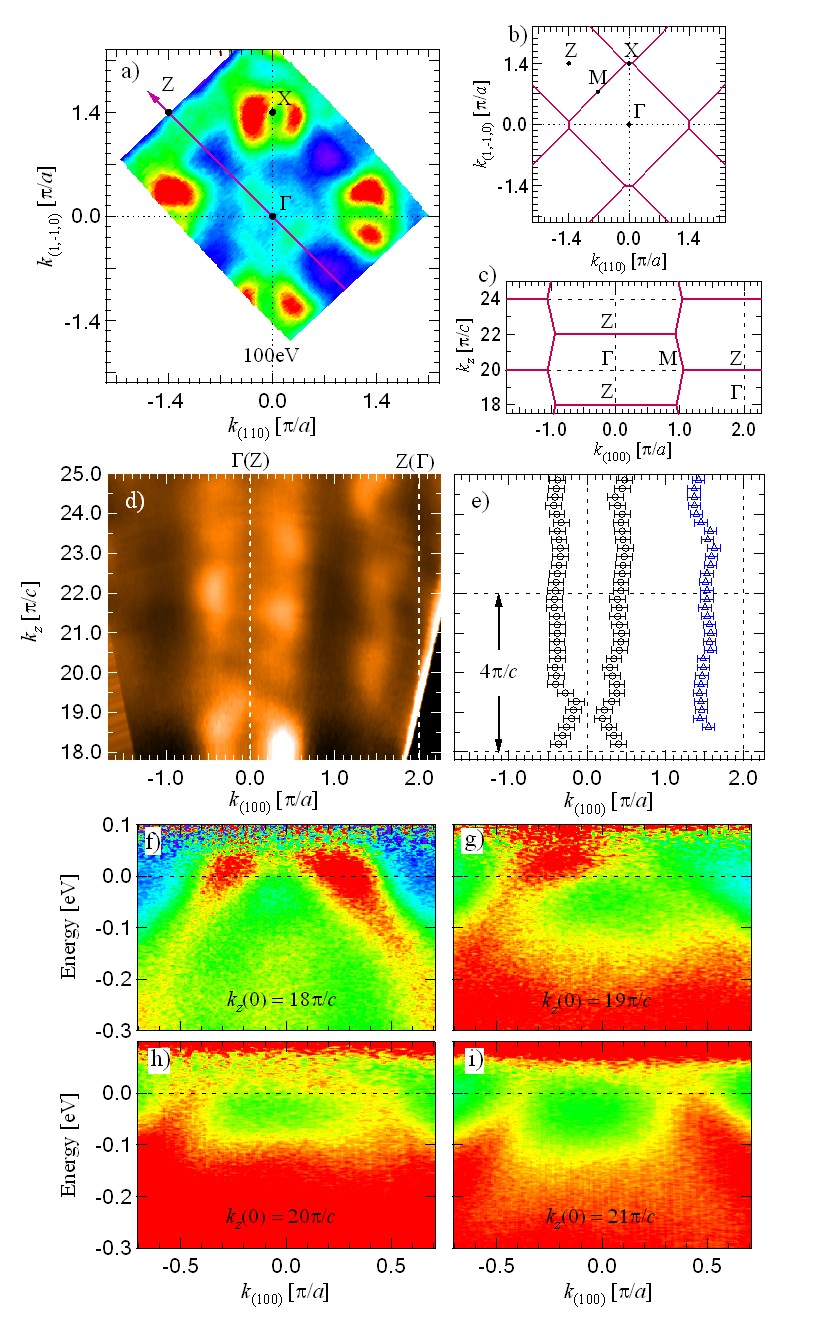}
\caption{(color online) $k_z$ dispersion data for CaFe$_2$As$_2$ in
the high temperature tetragonal phase ($T=200$K). (a) FS maps for
$h\nu=100$eV. (b) Brillouin zone structure in the $k_x$-$k_y$ plane
for the measured range. (c) Same as (b) but in the $k_x$-$k_z$ ($\Gamma$-Z) plane.
(d) $k_z$ dispersion data parallel to $\Gamma$-M (marked by magenta
arrow in panel (a)). The corresponding photon energy range was 80 to
190 eV. (e) Fermi crossing momenta $k_\textrm{F}$s extracted from
the data in panel (d). (f)-(i) Band dispersion data along a direction parallel to
$\Gamma$-M for $k_z=18$, 19, 20 and 21$\pi$ at $k_x=0$
(corresponding to incident photon energies of $h\nu=80, 89, 99$ and
110eV). The data was divided by the resolution convoluted Fermi
function to reveal the dispersion in the vicinity of $\mu$.} \label{fig3}
\end{figure}

\begin{figure}
\includegraphics[width=2.4in]{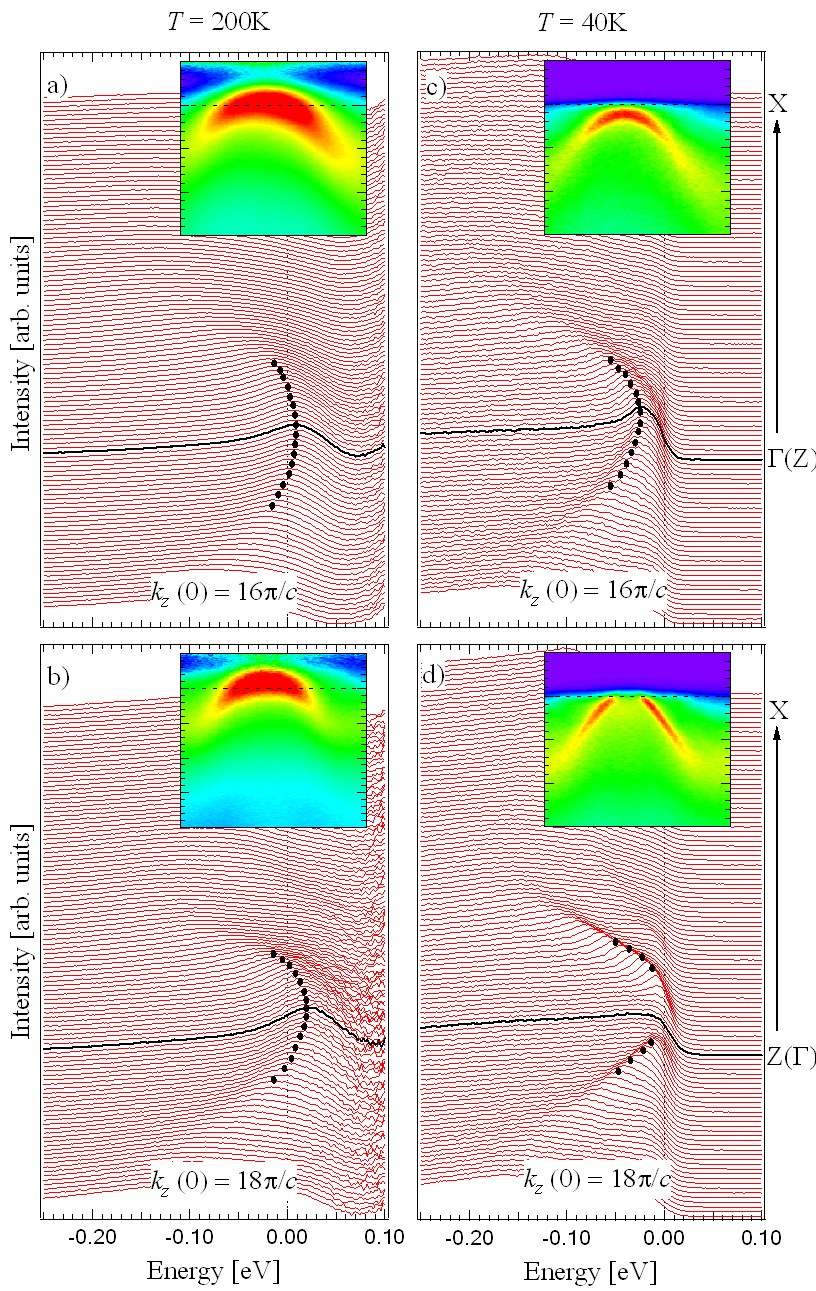}
\caption{(color online) Direct comparison of band dispersions for
low and high temperature phase of CaFe$_2$As$_2$. Each panel shows
the EDCs in vicinity of the $\Gamma$-point (black solid curve) along
$k_{(110)}$ direction. High temperature data was divided by the
resolution convoluted Fermi function to reveal dispersion in the
vicinity of $\mu$. Insets show the corresponding plot of band
dispersion. Data was obtained at temperatures and momenta indicated
in the panels.}
\end{figure}

Now we turn our attention to the high temperature \textsf{Tet}-phase
of CaFe$_2$As$_2$. We demonstrate in Fig. 3 the 2D character of the
band structure in the \textsf{Tet}-phase. Panel 3(a) shows the FS
map at $h\nu=100$eV. An arrow represents the $\Gamma$-Z direction
along which data in panels (d) and (e) were obtained. Schematic
arrangement of the Brillouin zones along in-plane and out-of-plane
directions for the \textsf{Tet}-phase are shown in panels 3(b) and
3(c). Panels 3(d) and 3(e) presents the actual $k_z$ intensity data
integrated within 10 meV about $\mu$ as a function
of $k_z$ and $k_x$ and the band dispersion extracted using MDC peaks
respectively. The photon energy range used here is 80 to 190eV,
which corresponds to $18\pi\leq k_z\leq25\pi$. The most important
observation here is that the bands around $\Gamma$ cross $\mu$ at
\textit{all} $k_z$s measured, no apparent $k_z$ dispersion is
visible. For further clarification, in panels 3(f)-3(i) we show
ARPES intensity maps divided by the resolution convoluted Fermi
function for $k_z=18,19,20$ and $21\pi$ respectively. The band
crosses $\mu$ for all these $k$-points, in clear contrast with the
situation for the low temperature \textsf{Ortho}-phase. The data in
Fig. 3 is consistent with a quasi-2D nature of the FS in the high
temperature \textsf{Tet}-phase of CaFe$_2$As$_2$.

In Fig. 4 we directly compare the band dispersion of low and high
temperature phase for the two $k_z$ values that correspond to high
symmetry points. The data is taken along $\Gamma$-X direction on the
same sample under exactly the same experimental conditions to avoid
possible complications due to scattering matrix elements or
polarization of incident photons. Though the data in Fig. 4 is taken
from different samples and beam lines than that in Fig. 2, the 3D
nature of the low temperature \textsf{Ortho}-phase reproduces
nicely. At high temperature we divided the data by the resolution
convoluted Fermi function to better see the location of the band in
the proximity of $\mu$. At low temperature this is
not necessary, as the width of the Fermi edge is much sharper than
the leading edge of the peaks, band crossings are clearly visible.
At $T=200$K, the $\alpha$-band crosses $\mu$ at both points
($k_z=16\pi$ and $k_z=18\pi$ - Figs. 4(a) and 4(b)). At low
temperature in the \textsf{Ortho}-phase, the same band crosses $\mu$
at $k_z=18\pi$ (Fig. 4(d)), but is located several tens of meV below
$\mu$ at $k_z=16\pi$ (Fig. 4(c)). This may be the
origin of the sudden drop in the electric resistivity found by
transport measurements when the material is heated above the
transition temperature \cite{NiNi_transition,Wu_FeAsflux}.

In conclusion, we have measured the in-plane and out-of-plane band
dispersion for both the orthorhombic (\textsf{Ortho}) and tetragonal
(\textsf{Tet}) phase of the iron arsenic A122 parent compound
CaFe$_2$As$_2$.
%%%%%%%
A number of theoretical models of the pairing mechanism and magnetic
ordering in these materials are based on nesting between different
sheets of the Fermi surface \cite{Mazin,Dong}. Our results
demonstrate that some FS sheets are indeed three dimensional,
therefore put significant constrains on possible nesting scenarios,
since the degree of nesting will strongly depend on the
dimensionality of the FSs.
%%%%%%%
Our finding also has important implications for understanding a
number of other physical properties such as the anisotropy in
electrical and thermal conductivity \cite{Makariy} which depend on
the dimensionality of the electronic structure.

We thank J. Schmalian, M. A. Tanatar and Rafael Fernandes for
insightful discussions and staff at SLS and ALS for excellent instrumentation support.
Ames Laboratory was supported by the Department of Energy - Basic
Energy Sciences under Contract No. DE-AC02-07CH11358. ALS is
operated by the US DOE under Contract No. DE-AC03-76SF00098.

\end{document}